\documentclass[aps,pra,twocolumn]{revtex4-1}
\usepackage{color}
\usepackage{array}
\usepackage{textcomp}
\usepackage{bm}
\usepackage{braket}
\usepackage{amsmath}
\usepackage{amssymb}
\usepackage{nccmath}
\usepackage{upgreek}
\usepackage{multirow}
\usepackage{graphicx}%\usepackage[draft]{graphicx}
\usepackage{hyperref}% add hypertext capabilities
\hypersetup{colorlinks=true, citecolor=blue, urlcolor=blue, linkcolor=blue}

\begin{document}
\title{%Field-Free
Long-Lasting Orientation of Symmetric-top Molecules \\ Excited by Two-Color Femtosecond Pulses} % of Chiral Molecules}

\author{Long Xu}
\thanks{These authors contributed equally to this work.}
\affiliation{AMOS and Department of Chemical and Biological Physics, The Weizmann Institute of Science, Rehovot 7610001, Israel}
\author{Ilia Tutunnikov}
\thanks{These authors contributed equally to this work.}
\affiliation{AMOS and Department of Chemical and Biological Physics, The Weizmann Institute of Science, Rehovot 7610001, Israel}
\author{Yehiam Prior}
\email{yehiam.prior@weizmann.ac.il}
\affiliation{AMOS and Department of Chemical and Biological Physics, The Weizmann Institute of Science, Rehovot 7610001, Israel}
\author{Ilya Sh. Averbukh}
\email{ilya.averbukh@weizmann.ac.il}
\affiliation{AMOS and Department of Chemical and Biological Physics, The Weizmann Institute of Science, Rehovot 7610001, Israel}

\begin{abstract}
Impulsive orientation of symmetric-top molecules excited by two-color femtosecond pulses is considered. In addition to the well-known transient orientation appearing immediately after the pulse and then reemerging periodically due to quantum revivals, we report the phenomenon of \emph{field-free long-lasting} orientation. Long-lasting means that the time averaged orientation remains non-zero until destroyed by other physical effects, e.g. intermolecular collisions.
The effect is caused by the combined action of the field-polarizability and field-hyperpolarizability interactions.
The dependence of degree of long-lasting orientation on temperature and pulse's parameters is considered. The effect can be measured by means of second (or higher-order) harmonic generation, and may be used to control the deflection of molecules traveling through inhomogeneous electrostatic fields.
\\
\\
\textbf{ Keywords: long-lasting orientation, symmetric-top, two-color laser pulses, polarizability interaction, hyperpolarizability interaction} %All article types: you may provide up to 8 keywords; at least 5 are mandatory.
\end{abstract}
\maketitle

\section{Introduction} \label{sec:Introduction}

Over the years, diverse optical methods have been developed to align and orient molecules of varying complexity  and many applications related to studies of molecular and photon-induced processes are based on the ability to control the absolute orientation of the molecules. For reviews, see \citep{StapelfeldtSeidman2003,Ohshima2010,Fleischer2012,Lemeshko2013,Koch2019Quantum,Lin2020Review}.

There are several laser-based strategies for achieving molecular orientation in the gas phase, including using a combination of intense non-resonant laser and weak electrostatic fields  \citep{Friedrich1999,Sakai2003,Goban2008,Ghafur2009,Holmegaard2009,Mun2014,Takei2016,Omiste2016,Thesing2017}, and using strong single-cycle terahertz (THz) pulses \citep{Harde1991,Averbukh2001,Machholm2001,Fleischer2011,Kitano2013,Damari2016,Babilotte2016Observation,Xu2020}, alone or together with optical pulses \citep{Daems2005,Gershnabel2006,Egodapitiya2014}. In addition, laser and THz pulses with twisted polarization were shown to be effective for inducing enantioselective orientation of chiral molecules \citep{Yachmenev2016,Gershnabel2018,Tutunnikov2018,Milner2019Controlled,Tutunnikov2019Laser,Tutunnikov2020Observation,TutunnikovXu2020}.

The techniques listed above rely on the laser-dipole and/or laser-polarizability interactions. Another route to molecular orientation stems from higher order laser-molecule interactions, e.g. the laser field - hyperpolarizability interaction. Non-resonant phase-locked two-color laser pulses consisting of the fundamental wave (FW) and its second harmonic (SH) were used for inducing molecular orientation by interacting with the molecular hyperpolarizability \citep{Vrakking1997,Dion1999,Kanai2001,Takemoto2008,De2009,Oda2010,JW2010,Zhang2011-multicolor,Frumker2012,Spanner2012,Znakovskaya2014,Mun2018,Mun2019,Mun2020,MelladoAlcedo2020,Shuo2020}.

Here, we investigate the orientation dynamics of symmetric-top molecules excited by single two-color femtosecond laser pulses. In addition to the well-known transient orientation appearing immediately as a response to the laser excitation, we predict the existence  of \textit{long-lasting orientation}. Long-lasting means that the time-averaged orientation remains non-zero, within the model, indefinitely or until destroyed by additional physical effects, e.g. by collisions. The long-lasting orientation induced by a two-color pulse has an intricate dependence on \emph{both} the molecular polarizability and hyperpolarizability. Related effects have been recently observed in chiral molecules excited by one-color laser pulses with twisted polarization \citep{Milner2019Controlled,Tutunnikov2020Observation} and investigated in non-linear molecules excited by THz pulses \citep{Xu2020,TutunnikovXu2020}.

The paper is organized as follows. In the next section, we describe our numerical approaches for simulating the laser-driven molecular rotational dynamics. In Sec. \ref{sec:SecII-long-lasting-orientation-effect}, we present the long-lasting orientation, which is the main result of this work. Section \ref{sec:SecIII-Qualitative-description} is devoted to a qualitative analysis of the effect, and a derivation of the approximate classical formula for the degree of long-lasting orientation. Additional results are presented in Sec. \ref{sec:SecV-additional-results}.

\section{Numerical methods} \label{sec:Numerical methods}

In this work, the rotational dynamics of symmetric-top molecules is treated within the rigid rotor approximation. We performed both classical and quantum mechanical simulations of molecular rotation driven by two-color laser fields.
This section outlines the theoretical approaches used in both cases.

\subsection{Classical simulation} \label{sec:classical simulation}
In the classical limit, the rotational dynamics of a single rigid top is described by Euler's equations \cite{Goldstein2002Classical}
\begin{equation}
\mathbf{I}\bm{\dot{\Omega}}=(\mathbf{I}\bm{\Omega})\times\bm{\Omega}+\mathbf{T},\label{eq:Eulers-equations}
\end{equation}
where $\mathbf{I}=\mathrm{diag}(I_{a},I_{b},I_{c})$ is the moment of inertia tensor, $\bm{\Omega}=(\Omega_{a},\Omega_{b},\Omega_{c})$ is the angular velocity, and $\mathbf{T}=(T_{a},T_{b},T_{c})$ is the external torque resulting from the interaction between field-induced dipole moment and the electric field. All the quantities in Eq. \eqref{eq:Eulers-equations} are expressed in the rotating molecular frame of reference, equipped with a basis set including the three principal axes of inertia, $a$, $b$, and $c$.

In the laboratory frame of reference, the electric field of a two-color laser pulse is defined by
\begin{equation}
\mathbf{E}(t) = \varepsilon_1(t)\cos(\omega t)\bm{e}_Z +  \varepsilon_{2}(t)\cos(2\omega t+\varphi)\bm{e}_{\mathrm{SH}}
,\label{eq:electric-field-lab-frame}
\end{equation}
where the two terms correspond to the FW and its SH, respectively. $\omega$ is the carrier frequency of the FW field, $\varphi$ is the relative phase of the second harmonic,
$\varepsilon_{n}(t)= \varepsilon_{n,0}\exp[-2\ln 2\, (t/\sigma_n)^2],\, n=1,2, $ is the field's envelope with $\varepsilon_{n,0}$ as the peak amplitude, and $\sigma_n$ is the full width at half maximum (FWHM) of the laser pulse intensity profile.
The polarization direction of the SH field is given by
$\bm{e}_{\mathrm{SH}} = \cos(\phi_{\mathrm{SH}})\bm{e}_Z + \sin(\phi_{\mathrm{SH}})\bm{e}_X$,
where $\phi_{\mathrm{SH}}$ is its angle with respect to the $Z$ axis,
$\bm{e}_{Z}$ and $\bm{e}_{X}$ are the unit vectors along laboratory $Z$ and $X$ axes, respectively.
The electric field in the molecular frame of reference can be expressed as
\begin{equation}
\mathbf{E}(t) = \varepsilon_1(t)\cos(\omega t)\bm{e}_1 +  \varepsilon_{2}(t)\cos(2\omega t+\varphi)\bm{e}_{2}, \label{eq:electric-field-molecular-frame}
\end{equation}
where $\bm{e}_{1} = Q \bm{e}_{Z}$ and $\bm{e}_{2} = Q \bm{e}_{\mathrm{SH}}$ are the unit vectors expressed in the molecular frame of reference.
$Q$ is a $3 \times 3$ time-dependent orthogonal matrix relating the laboratory and the molecular frames of reference. It is parametrized by a quaternion, $q$ which has an equation of motion $\dot{q}=q\Omega/2$, with $\Omega=(0,\bm{\Omega})$ being a pure quaternion \citep{Coutsias2004The,Kuipers1999Quaternions}.
Considering laser-polarizability and laser-hyperpolarizabiltiy interactions, the torque induced by a two-color field has two contributions $\mathbf{T} = \mathbf{T}^\alpha + \mathbf{T}^\beta$, where  \cite{Lin2018All}
\begin{align}
\mathbf{T}^\alpha &= \overline{(\bm{\alpha}\mathbf{E}\times \mathbf{E})}
= \frac{\varepsilon_1^2}{2}(\bm{\alpha}\bm{e}_1)\times \bm{e}_1
+ \frac{\varepsilon_2^2}{2}(\bm{\alpha}\bm{e}_2)\times \bm{e}_2, \label{eq:average-torque-alpha} \\
\mathbf{T}^\beta_i &=\frac{1}{2}\overline{[\left(\mathbf{E}\bm{\beta} \mathbf{E}\right)\times \mathbf{E}]}_i = \sum\limits_{m,n,j,k} \frac{\varepsilon_1^2 \varepsilon_2}{4}\epsilon_{ijk}
\beta_{m n j} e_{1 m} e_{2 n} e_{1 k}\nonumber\\
&+\sum\limits_{m,n,j,k}\frac{\varepsilon_1^2 \varepsilon_2}{8}\epsilon_{ijk} \beta_{m n j} e_{1 m} e_{1 n} e_{2 k}.\label{eq:average-torque-beta}
\end{align}
Here, the overline $\overline{(\cdots)}$ represents averaging over the optical cycle, $\bm{\alpha}$ and $\bm{\beta}$ are the polarizability and hyperpolarizability tensors, respectively.
$\epsilon_{ijk}$ is the Levi-Civita symbol, $\beta_{mnj}$ is a component of the hyperpolarizability tensor,
$e_{1m}$ and $e_{2m}$ are the components of the FW and SH fields, respectively.

To simulate the behavior of an ensemble of non-interacting molecules, we use the Monte Carlo approach. For each molecule, the Euler's equations [Eq. \eqref{eq:Eulers-equations}] with the torques in Eqs. \eqref{eq:average-torque-alpha} and \eqref{eq:average-torque-beta} are solved numerically using the standard fourth order Runge-Kutta algorithm.
In the simulations we used ensembles consisting of $N\gg1$ molecules.
The initial uniform random quaternions, representing isotropically distributed molecules, were generated using the recipe from \citep{Lavalle2006Planning}.
Initial angular velocities are distributed according to the Boltzmann distribution,
\begin{equation}
f(\bm{\Omega})\propto %\exp\left(-\frac{\bm{\Omega}^{T}\mathbf{I}\bm{\Omega}}{2k_{B}T}\right)=
\prod_{i}\exp\left(-\frac{I_{i}\Omega_{i}^{2}}{2k_{B}T}\right),\label{eq:Classical-Boltzmann-distribution}
\end{equation}
where $i=a,b,c$, $T$ is the temperature and $k_{B}$ is the Boltzmann constant.

\subsection{Quantum simulation} \label{sec:quantum simulation}

The Hamiltonian describing the rotational degrees of freedom of a molecule and the molecular polarizability and hyperpolarizability couplings to external time-dependent fields can be written  as
$H(t)=H_{r} + H_\mathrm{int}(t)$, where $H_{r}$ is the field-free Hamiltonian \citep{zare1988Angular}, and  $H_\mathrm{int}(t)$ is the  molecule-field interaction potential, with two contributions
$H_\mathrm{int}(t) = V_{\alpha} + V_{\beta} $, where \citep{Buckingham2007Permanent}
\begin{equation}\label{eq:potential}
V_{\alpha} = -\frac{1}{2} \sum_{i, j} \alpha_{i j} E_{i} E_{j}, \quad
V_{\beta} = -\frac{1}{6} \sum_{i, j, k} \beta_{i j k} E_{i} E_{j} E_{k}.
\end{equation}
Here $E_{i}$, $\alpha_{i j}$, and $\beta_{ijk}$ are the components of the field vector, polarizability tensor $\bm{\alpha}$, and hyperpolarizability tensor $\bm{\beta}$, respectively.
Since the optical carrier frequency of the laser fields, $\omega$ [see Eq. \eqref{eq:electric-field-lab-frame}], is several orders of magnitude larger than a typical rotational frequency of small molecules, the energy contribution due to the interaction with the molecular permanent dipole, $\bm{\mu}$, $-\bm{\mu} \cdot \mathbf{E}(t)$ is negligible.

\begin{table*}[t]
\begin{centering}
\begin{tabular}{c|c|c|c}
Moments of inertia & Dipole components & Polarizability components & Hyperpolarizability components\tabularnewline
\hline
$I_{a}=20982$ & $\mu_{a}=-0.736$ & $\alpha_{aa}=18.38$ & $\beta_{aaa}=40.449$\tabularnewline
$I_{b}=129238$ &  & $\alpha_{bb}=16.76$ & $\beta_{abb}=\beta_{acc}=26.970$\tabularnewline
$I_{c}=129238$ &  & $\alpha_{cc}=16.76$ & $\beta_{bbb}=-\beta_{bcc}=-11.019$\tabularnewline
\end{tabular}
\par\end{centering}
\caption{Molecular properties (in atomic units) of $\mathrm{CH_{3}F}$: moments
of inertia, nonzero elements of dipole moment, polarizability tensor,
and hyperpolarizability tensor. All the quantities are represented in the reference frame of molecular principal axes of inertia.
\label{tab:Molecular-properties-alternative}}
\end{table*}

We use the eigenstates of $H_r$, $|JKM\rangle$, describing the field-free motion of quantum symmetric-top \citep{zare1988Angular}, as the basis set in our numerical simulations. The three quantum numbers are $J$, $K$ and $M$, where $J$ is the total angular momentum, while $K$ and $M$ are its projections on the molecular $a$ axis and the laboratory-fixed $Z$ axis, respectively. The time-dependent Schr\"{o}dinger equation $i\hbar \partial_t |\Psi(t)\rangle = H(t)|\Psi(t)\rangle$ is solved by numerical exponentiation of the Hamiltonian matrix (see Expokit \cite{sidje1998Expokit}) with the initial state being one of the field-free eigenstates, $|\Psi(t=0)\rangle=|J K M\rangle$.
The degree of molecular orientation is derived by calculating the induced polarization, the expectation value of the dipole projection.
The polarization along each of the axes in the laboratory-fixed frame of reference is given by
\begin{align}\label{eq:Polarization}
& \braket{\mu_i^{(J,K,M)}}\!(t)  =\langle\Psi(t)|\bm{\mu}\cdot\mathbf{e}_{i}|\Psi(t)\rangle,
\end{align}
where $\mathbf{e}_{i}$ represents one of the unit vectors $\mathbf{e}_{X}, \mathbf{e}_{Y}, \mathbf{e}_{Z}$.
Thermal effects are accounted for by computing the incoherent average of the time-dependent polarizations obtained for the various initial states $|J K M\rangle$. The relative weight of each of the projections $\braket{\mu_i^{(J,K,M)}}\!(t)$ is defined by the Boltzmann distribution,
\begin{flalign}
\braket{\mu_i}\!(t) =\frac{1}{\mathcal{Z}}\sum\limits _{J,K,M}\epsilon_{K}
%e^{-\frac{\varepsilon_{J,K,M}}{k_{B}T}}
\exp\left[-\frac{\varepsilon_{J,K,M}}{k_{B}T}\right]
\braket{\mu_i^{(J,K,M)}}\!(t),\label{eq:Quantum-Boltzmann-distribution-Symmetrictop}
\end{flalign}
where $\mathcal{Z}=\sum_{J,K,M}\epsilon_{K}\exp\left(-\varepsilon_{J,K,M}/k_{B}T\right)$ is the partiton function, and $\varepsilon_{J,K,M}$ is the energy/eigenvalue corresponding to $|J K M\rangle$ state.
For molecules with two or more identical atoms, an additional statistical factor $\epsilon_{K}$ must be included in the distribution \citep{McDowell1990Rotational}.
For the case of methyl fluoride ($\mathrm{CH_{3}F}$) molecule considered in this work, $\epsilon_{K}$ is given by
\begin{equation}
\epsilon_{K}=\frac{(2I_{H}+1)^{3}}{3}\left[1+\frac{2\cos(2\pi K/3)}{(2I_{H}+1)^{2}}\right],\label{eq:statistic-weight}
\end{equation}
where $I_{H}=1/2$.

In our simulations, the basis set included all the states with $J\leq30$. For our sample molecule, $\mathrm{CH_{3}F}$, at initial  temperature of $T=5\, \mathrm{K}$, this means that initial states with $J\leq8$ were included. Additional details about the numerical simulations, including the matrix elements of the interaction Hamiltonian $H_\mathrm{int}$, can be found in Appendix \ref{app:quantum simulation}.

\section{Long-lasting orientation}  \label{sec:SecII-long-lasting-orientation-effect}

We continue to consider the methyl fluoride ($\mathrm{CH_{3}F}$), as an example for  a symmetric-top molecule. The molecule is excited by a two-color pulse in which the polarizations of the FW and SH are parallel and along $Z$ axis [$\phi_\mathrm{SH}=0$, see Eq. \eqref{eq:electric-field-lab-frame}]. In addition, here we set the relative phase between them to be zero ($\varphi=0$). Later on we discuss what changes when this phase changes.
Table \ref{tab:Molecular-properties-alternative} summarizes the molecular properties of $\mathrm{CH_{3}F}$.
Moments of inertia, dipole moment, and polarizability tensor components are taken from NIST, where they were computed within the density functional theory (DFT, method CAM-B3LYP/aug-cc-pVTZ) \citep{johnson1999nist}. The hyperpolarizability values are literature values taken from \citep{Chong1992}.

\begin{figure}[!b]
\begin{centering}
\includegraphics[width=\linewidth]{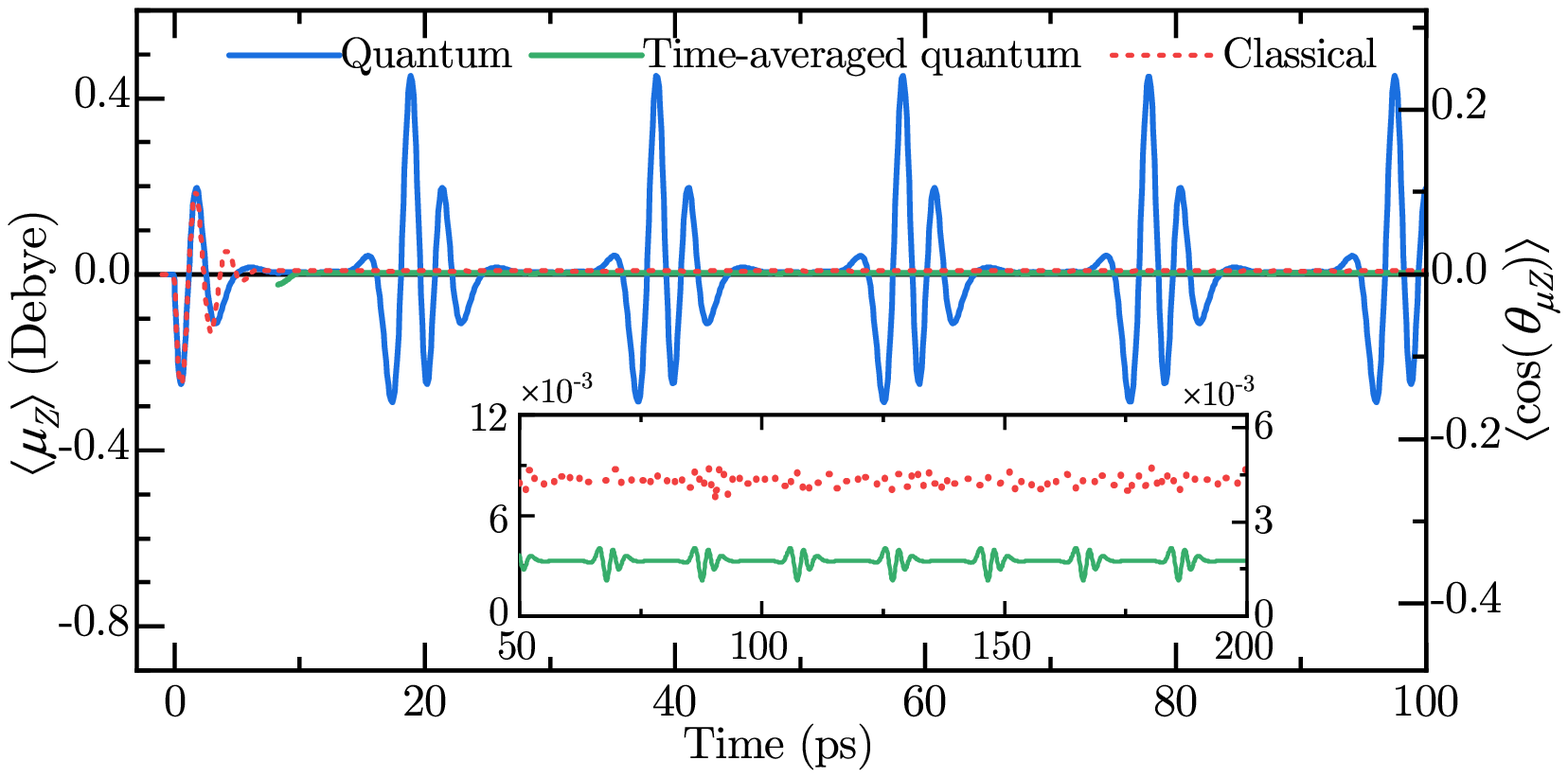}
\par\end{centering}
\caption{ $Z$-projection
of the dipole moment, $\braket{\mu_{Z}}$ and the orientation factor, $\braket{\cos(\theta_{\mu Z})}\equiv \braket{\mu_{Z}}/\mu$ as a function of time for
$\mathrm{CH_{3}F}$ molecule at initial rotational temperature $T=5\,\mathrm{K}$.
Here $\mu$ is the magnitude of the dipole moment and $\theta_{\mu Z}$ denotes the angle between the dipole moment and laboratory
$Z$ axis. The solid blue and dotted red lines represent the results
of quantum and classical simulations, respectively. The solid green
line is the time average defined by $\overline{\langle\mu_{Z}\rangle(t)}=(\Delta t)^{-1}\int_{t-\Delta t/2}^{t+\Delta t/2}\mathrm{d}t'\langle\mu_{Z}\rangle(t')$,
where $\Delta t=19.6\,\mathrm{ps}$. The inset shows a magnified portion
of the signals. \label{fig:Fig1}}
\end{figure}

Figure \ref{fig:Fig1} shows the projection of the dipole moment along
the laboratory $Z$ axis, $\braket{\mu_{Z}}$, calculated classically and quantum mechanically
(see Methods Section \ref{sec:Numerical methods}). In the classical case, the angle brackets
$\braket{\cdots}$ denote ensemble average, that is the average of the dipole projections of $N=10^8$ molecules, initially isotropically distributed in space and having random angular velocities [see Eq. \eqref{eq:Classical-Boltzmann-distribution}]. In the quantum case, $\braket{\cdots}$ denotes incoherent average of initially populated rotational states  [see Eq. \eqref{eq:Quantum-Boltzmann-distribution-Symmetrictop}].
Note that the averages $\braket{\mu_{X}}$ and $\braket{\mu_{Y}}$ are zero.
Here, the initial temperature is $T=5\,\mathrm{K}$, the peak intensities of the FW
and SH fields are $I_{\mathrm{FW}}=8\times10^{13}\,\mathrm{W/cm^{2}}$
and $I_{\mathrm{SH}}=3\times10^{13}\,\mathrm{W/cm^{2}}$, respectively,
and the duration (FWHM) of the pulses are $\sigma_1=\sigma_2=120\,\mathrm{fs}$
[see Eq. \eqref{eq:electric-field-lab-frame}]. On the short
time scale (first $\approx2\,\mathrm{ps}$), the classical and quantum
results are in remarkable agreement, and show the expected immediate
response to a kick by a two-color pulse.
On the long time scale, the quantum mechanical simulation exhibits
distinct quantum revivals of the orientation \citep{Eberly1980,Parker1986,Averbukh1989,Felker1992Rotational,Robinett2004}. This transient orientation effect is well studied and was observed in the past \citep{Vrakking1997,Dion1999,Kanai2001,Takemoto2008,De2009,Oda2010,JW2010,Zhang2011-multicolor,Frumker2012,Spanner2012,Znakovskaya2014,Mun2018,Mun2019,Mun2020,MelladoAlcedo2020,Shuo2020}.

In the case of symmetric-top molecules considered
here, we observe \emph{long-lasting (persistent)} orientation, a previously unreported phenomenon in two-color orientation schemes. The inset in Fig. \ref{fig:Fig1} demonstrates  that after the initial oscillations are washed out, the classical polarization/degree of orientation attains a constant, nonzero value. In the quantum case too, despite its being partially masked by the revivals, the sliding time average of the signal is approximately constant and it persists indefinitely within the adopted model. This long-lasting orientation is one of the main results of this work.

Several comments are in order. Additional physical effects can distort the long-term field-free picture of identical periodically appearing revivals seen in Fig. \ref{fig:Fig1}. These include the centrifugal
distortion and the radiation emission due to rapidly rotating molecular
permanent dipole moment. Dephasing of the rotational states caused by the centrifugal distortion leads to the eventual decay of the revivals' peaks \citep{Babilotte2016Observation,Damari2016}. Nevertheless, the average dipole
remains almost unchanged (see \citep{Xu2020}).
The radiative emission results in the gradual decrease of the
rotational energy \citep{Babilotte2016Observation,Damari2016}.
However, for a rarefied molecular gas, the estimated relative
energy loss during a single revival is very small.
The proper description of the behavior on an even longer timescale (nanoseconds), requires the inclusion of collisions and fine structure effects \citep{Esben2018,Thesing2020}, which is beyond the scope of the current work. Furthermore, it should be noted that higher laser pulse intensities lead to higher degree of orientation, but when the intensity is high enough for molecular ionization, another effect kicks in, namely  orientation mechanism due to selective molecular ionization of molecules with specific orientation  \citep{Spanner2012,Znakovskaya2014}. Considering this kind of orientation is also beyond the scope of this work.

\section{Long-lasting orientation - a qualitative description} \label{sec:SecIII-Qualitative-description}

An explicit form of the interaction potential [Eq. \eqref{eq:potential}] can be obtained
by expressing the electric field vector in the rotating molecular frame of reference.
For the  sake of the current discussion, this can be done conveniently by using an orthogonal rotation matrix
parameterized by the three Euler angles, $R(\phi,\theta,\chi)$. We
use the definition convention adopted in \citep{zare1988Angular}, according to which, $\phi$ and
$\theta$ are the standard azimuth and polar angles defining the orientation
of the molecular frame $z$ axis, and $\chi$ is the additional rotation angle
about $z$ axis. The basis set in the rotating molecular frame of reference consists of the three principal
axes of inertia, $a,b,c$. For molecules belonging to the $C_{3v}$ symmetry group
(such as $\mathrm{CH_3F}$), there are three non-zero polarizability
components (two of them are equal), and 11 non-zero hyperpolarizability components (three of which are independent) \citep{Buckingham1967}.
For definiteness, we associate the axis of the three-fold rotational
symmetry with the most polarizable molecular principal axis $a$ ($z$
axis in the rotating frame), having the smallest moment of inertia,
$I_{a}$. In this case, the non-zero polarizability elements are $\alpha_{aa}>\alpha_{bb}=\alpha_{cc}$,
and the independent hyperpolarizability elements are $\beta_{aaa}$, $\beta_{abb}=\beta_{acc}$, $\beta_{bbb}=-\beta_{bcc}$. The other non-zero hyperpolarizability elements are obtained by permuting the indices of the independent elements \citep{Buckingham1967}. The parameters of the $\mathrm{CH_3F}$ molecule are listed in Table \ref{tab:Molecular-properties-alternative}.

\begin{figure}
\begin{centering}
\includegraphics[width=\linewidth]{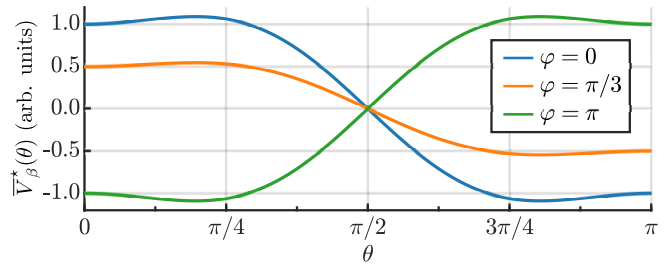}
\par\end{centering}
\caption{Angle dependence of the potential $\overline{V}_{\beta}^\star(\theta)$,
$\cos(\varphi)[3b\sin^{2}(\theta)\cos(\theta)+\cos^{3}(\theta)]$, see Eq. \eqref{eq:potential-hyperpolarizability2},
for different relative phases, $\varphi$. Here, $b=\beta_{abb}/\beta_{aaa}=2/3$.
\label{fig:Fig2}}
\end{figure}

We consider the case of a two-color pulse in which both the FW and SH are polarized along $Z$ axis. The interaction potential is obtained by carrying out the summation in Eq. \eqref{eq:potential} (where all the quantities are expressed in the basis of principal axes of  inertia) and we average over the optical cycle. The resulting potential has two contributions,
\begin{small}
\begin{align}
\overline{V}_{\alpha}(\theta) & =-\frac{\varepsilon_{1}^{2}(t)+\varepsilon_{2}^{2}(t)}{4}(\alpha_{aa}-\alpha_{bb})\cos^{2}(\theta),\label{eq:potential-polarizability}\\
\overline{V}_{\beta}(\theta,\chi) & =-\frac{\varepsilon_{1}^{2}(t)\varepsilon_{2}(t)}{8}\cos(\varphi)\sin^{3}(\theta)\cos(3\chi)\beta_{bcc}\nonumber \\
-\frac{\varepsilon_{1}^{2}(t)\varepsilon_{2}(t)}{8} & \cos(\varphi)\left[3\sin^{2}(\theta)\cos(\theta)\beta_{abb}+\cos^{3}(\theta)\beta_{aaa}\right].\!
\label{eq:potential-hyperpolarizability}
\end{align}
\end{small}
To facilitate the qualitative discussion in this section, we let $\beta_{bbb}=\beta_{bcc}=0$.
These elements of the hyperpolarizability tensor are the smallest (see Table \ref{tab:Molecular-properties-alternative}), and
their omission does not affect the qualitative features of the discussed phenomena.
Thus, the hyperpolarizability interaction becomes
\begin{align}
\overline{V}_{\beta}^\star(\theta) & =
-\frac{\varepsilon_{1}^{2}(t)\varepsilon_{2}(t)}{8}  \cos(\varphi) \nonumber\\
&\times\left[3\sin^{2}(\theta)\cos(\theta)\beta_{abb}+\cos^{3}(\theta)\beta_{aaa}\right].\!
\label{eq:potential-hyperpolarizability2}
\end{align}
And  $\overline{V}=\overline{V}_\alpha+\overline{V}_\beta^\star$ is a function of a single variable $\theta$---the polar angle between the symmetry axis of the molecule ($a$ axis)
and the laboratory $Z$ axis (axis of laser polarization).

The two parts of the interaction potential lead to
two distinct effects. $\overline{V}_{\alpha}(\theta)$
is a symmetric function of $\theta$ (about $\theta=\pi/2$), and
a kick by such a potential results in molecular alignment (for reviews, see \citep{StapelfeldtSeidman2003,Ohshima2010,Fleischer2012,Lemeshko2013,Koch2019Quantum,Lin2020Review}). The second part, $\overline{V}_{\beta}^\star(\theta)$ is an asymmetric
function of $\theta$, causing molecular orientation. Transient
orientation of linear molecules excited by two-color laser pulses
has been observed \citep{De2009,Oda2010,JW2010,Frumker2012,Znakovskaya2014} and is being  studied theoretically \citep{Mun2018,Mun2019,Mun2020,MelladoAlcedo2020,Shuo2020}.
Figure \ref{fig:Fig2} shows the angular dependence of $\overline{V}_{\beta}^\star(\theta)$, $\cos(\varphi)[3b\sin^{2}(\theta)\cos(\theta)+\cos^{3}(\theta)]$, see Eq. (\ref{eq:potential-hyperpolarizability2}).
The orienting potential is proportional to $\cos(\varphi)$, such that the orientation is zero for $\varphi=\pi/2$. Also, the relative phase can be used to control the orientation direction. To simplify the following expressions, we set $\varphi=0$.

\subsection{Approximate classical formula}

\begin{figure}
\begin{centering}
\includegraphics[width=6cm]{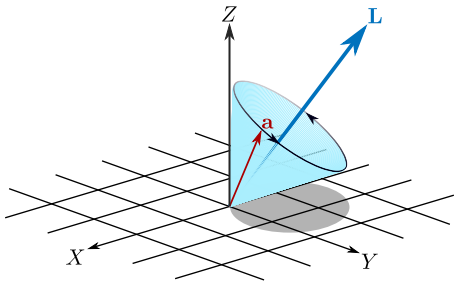}
\par\end{centering}
\caption{Illustration of precession of the vector $\mathbf{a}$ about the angular momentum vector $\mathbf{L}$. The tip of $\mathbf{a}$ describes a circle, while the arrow lies on the surface of a cone. The angular frequency of the precession is $L/I$, see Eq. \eqref{eq:a(t)-symmetric}.
\label{fig:Fig3}}
\end{figure}

In the case of weak excitation, we can derive an approximate classical formula for the
degree of long-lasting orientation. Since the long-lasting orientation manifests itself under field-free conditions, we begin by considering the free motion of a single classical symmetric top.
The free motion of the unit vector $\mathbf{a}$,
pointing along the rotational symmetry axis of the molecule, is given
by a simple vectorial differential equation $\dot{\mathbf{a}}=(\mathbf{L}/I)\times\mathbf{a}$.
Here, $\mathbf{L}$ is the conserved angular momentum vector,
$I$ is the moment of inertia along the orthogonal axes $b$ and $c$ $(I_{a}<I_{b}=I_{c}\equiv I)$.
The solution of this equation is given by
\begin{align}
\mathbf{a}(t) & =\mathbf{L}\frac{\mathbf{L}\cdot\mathbf{a}(0)}{L^{2}}\nonumber +\left[\mathbf{a}(0)-\mathbf{L}\frac{\mathbf{L}\cdot\mathbf{a}(0)}{L^{2}}\right]\cos\left(\frac{L}{I}t\right)\nonumber \\
 & +\frac{\mathbf{L}}{L}\times\mathbf{a}(0)\sin\left(\frac{L}{I}t\right),\label{eq:a(t)-symmetric}
\end{align}
where $L$ is the magnitude of angular momentum and $\mathbf{a}(0)$ is vector $\mathbf{a}$ at $t=0$.
The above equation describes precession of $\mathbf{a}$
around $\mathbf{L}$ at a rate $L/I$ (see Fig. \ref{fig:Fig3}). In the special case of a linear molecule, $L_{a}=\mathbf{L}\cdot\mathbf{a}(0)=0$, so that Eq. \eqref{eq:a(t)-symmetric} reduces to
\begin{equation}
\mathbf{a}(t)=\mathbf{a}(0)\cos\left(\frac{L}{I}t\right)+\frac{\mathbf{L}}{L}\times\mathbf{a}(0)\sin\left(\frac{L}{I}t\right).\label{eq:a(t)-linear}
\end{equation}
Equation \eqref{eq:a(t)-linear} describes a uniform rotation of $\mathbf{a}$ in a plane perpendicular
to the angular momentum vector $\mathbf{L}$.

The degree of long-lasting orientation (see the inset of Fig. \ref{fig:Fig1})
can be obtained by considering the ensemble average projection of the molecular axis $\mathbf{a}$ on the laboratory $Z$ axis, $a_Z=\mathbf{e}_Z\cdot\mathbf{a}$, and then evaluating its time average.
\begin{equation}
\overline{\braket{a_{Z}}}=\lim_{\tau\rightarrow\infty}\frac{1}{\tau}\int_{0}^{\tau}\braket{\mathbf{e}_Z\cdot\mathbf{a}(t)}\,dt,\label{eq:aZ-averages}
\end{equation}
where $t=0$ defines the end of the two-color pulse (when
the free motion begins). Note that for $\mathrm{CH_3F}$ the molecular dipole, $\bm{\mu}$ points along $-\mathbf{a}$ (see Table \ref{tab:Molecular-properties-alternative}).
Next, we exchange the order of the ensemble and time averaging. The time average of $a_{Z}$ is obtained from Eq. \eqref{eq:a(t)-symmetric} and it reads
\begin{align}
\overline{a_{Z}} & =\frac{(L_{Z})_{f}(L_{a})_{f}}{L_{f}^{2}},\label{eq:abar-f}
\end{align}
where $L_Z=\mathbf{e}_Z\cdot\mathbf{L}$, $L_a=\mathbf{a}\cdot\mathbf{L}$, and subindex $f$ denotes that all the quantities are taken
after the pulse. With the  potential in Eqs. (\ref{eq:potential-polarizability})
and (\ref{eq:potential-hyperpolarizability2}),
both $\phi$ and $\chi$ are cyclic coordinates. Therefore, the canonically
conjugate angular momenta $L_{Z}$ and $L_{a}$  are conserved.
As a consequence, Eq. (\ref{eq:abar-f}) becomes
\begin{equation}
\overline{a_{Z}}=\frac{L_{Z}L_{a}}{L_{f}^{2}},\label{eq:abar-i}
\end{equation}
where $L_{Z}$ and $L_{a}$ are taken before the pulse. At this stage, we can conclude that the long-lasting
orientation is strictly zero when the initial temperature is zero
and/or in the limit of a linear rotor. In the first
case, $L_{Z}=L_{a}=0$, while in the second case $L_{a}=0$, because $I_a=0$ for linear molecules.

For the ensemble averaging, it is advantageous to express
all the quantities in the basis of principal axes of inertia.
The magnitude of the angular momentum after the pulse, $L_{f}^{2}$,
is given by
\begin{align}
L_{f}^{2} & =(L_{b}+\delta L_{b})^{2}+(L_{c}+\delta L_{c})^{2}+L_{a}^{2},\label{eq:Lnew}
\end{align}
where $L_{a},L_{b},L_{c}$ are the values before the pulse, while $\delta L_{b}$
and $\delta L_{c}$ are the changes in angular momentum components due to laser excitation. Explicit expressions for $\delta L_{b}$
and $\delta L_{c}$ can be obtained using the impulsive approximation.
In this approximation, we assume that the duration of the two-color
pulse is much shorter than the typical period of molecular rotation,
such that the molecular orientation remains unchanged during
the pulse. Using this approximation and the Euler-Lagrange equations, we derive the explicit expressions for $\delta L_{b}$
and $\delta L_{c}$ (the details are summarized in Appendix \ref{app:torques}),
\begin{align}
\delta L_{b} & =f(\theta)\sin(\chi),\label{eq:delta-Lb}\\
\delta L_{c} & =f(\theta)\cos(\chi),\label{eq:delta-Lc}
\end{align}
where
\begin{align}
&f(\theta)  =P_{1}\sin(2\theta)\nonumber \\
+&P_{2}\sin(\theta)  [(3\cos(2\theta)+1)\beta_{abb}-2\cos^{2}(\theta)\beta_{aaa}],\label{eq:text-f(theta)}
\end{align}
and
\begin{align}
P_{1} & =\frac{\sigma}{4}\sqrt{\frac{\pi}{\ln(16)}}(\varepsilon_{1,0}^{2}+\varepsilon_{2,0}^{2})(\alpha_{bb}-\alpha_{aa})\label{eq:text-P1},\\
P_{2} & =\frac{3\sigma}{16}\sqrt{\frac{\pi}{\ln(64)}}\varepsilon_{1,0}^{2}\varepsilon_{2,0}.\label{eq:text-P2}
\end{align}
Here $\sigma=\sigma_1=\sigma_2$, $\varepsilon_{1,0}$ and $\varepsilon_{2,0}$ are the peak amplitudes of the FW and SH, respectively. $L_{Z}$ is expressed in terms of the molecular frame components
$L_{a,b,c}$ using the rotation matrix $R(\phi,\theta,\chi)$, such that
\begin{fleqn}\begin{equation}
L_{Z}=-\sin(\theta)\cos(\chi)L_{b}+\sin(\theta)\sin(\chi)L_{c}
+L_{a}\cos(\theta).\label{eq:LZ-in-terms-Labc}
\end{equation}\end{fleqn}
Finally, we carry out the ensemble average
\begin{align}
\langle\overline{a_{Z}}\rangle & \nonumber =\frac{1}{\mathcal{Z}}\int_\Omega\int_{L^3}\frac{L_{Z}L_{a}}{L_{f}^{2}}\exp\left[-\frac{1}{2k_{B}T}\left(\frac{L_{a}^{2}}{I_{a}}+\frac{L_{b}^{2}}{I}+\frac{L_{c}^{2}}{I}\right)\right]\nonumber \\
 & \times\sin\theta\,d\theta d\chi d\phi dL_{a}dL_{b}dL_{c},\label{eq:aZ-integral-1}
\end{align}
where $\mathcal{Z}$ is the partition function. To simplify the integral, we assume that $|\delta L_{b}/L_{b}|,|\delta L_{c}/L_{c}|\ll1$ [see Eqs. (\ref{eq:Lnew}-\ref{eq:delta-Lc})]
and expand $1/L_{f}^{2}$ in powers of $f(\theta)$ [see Eq. \eqref{eq:text-f(theta)}]. Only terms proportional
to even powers of $f(\theta)$ contribute to the integral. We consider
the first non-vanishing term proportional to $f^{2}(\theta)$, such
that [see Eq. \eqref{eq:App-aZ-integral-3}]
\begin{equation}
\langle\overline{a_{Z}}\rangle\approx\frac{\tilde{I}_{L}(w)}{4k_{B}TI}\sqrt{\frac{w}{\pi}}\int f^{2}(\theta)\sin(2\theta)\,d\theta,\label{eq:aZ-integral-2}
\end{equation}
where $w=I/I_{a}$, and $\tilde{I}_{L}(w)$ is a monotonic
function of $w>1$. In the limit of a linear molecule ($w\rightarrow\infty$), $\tilde{I}_{L}(w)\rightarrow0$ [see Eq. \eqref{eq:App-IL-final} and Fig. \ref{fig:app-Fig1}].

For the polarizability interaction alone,
$f(\theta)=P_{1}\sin(2\theta)$, and $\int f^{2n}(\theta)\sin(2\theta)\,d\theta=0$.
For the hyperpolarizability interaction alone, $f(\theta)=P_{2}\sin(\theta)[(3\cos(2\theta)+1)\beta_{abb}-2\cos^{2}(\theta)\beta_{aaa}]$
which is a symmetric function (about $\theta=\pi/2$), and therefore $\int f^{n}(\theta)\sin(2\theta)\,d\theta=0$
for all $n$. Only when both polarizability and hyperpolarizability
interactions are included, $\langle\overline{a_{Z}}\rangle\neq0$.
In this case,
\begin{equation}
\langle\overline{a_{Z}}\rangle\approx\frac{16\tilde{I}_{L}(w)}{105k_{B}TI}\sqrt{\frac{w}{\pi}}P_{1}P_{2}\left(2\beta_{abb}-3\beta_{aaa}\right),\label{eq:aZ-final}
\end{equation}
where $P_{1}$ and $P_{2}$ are given by Eqs. (\ref{eq:text-P1})
and (\ref{eq:text-P2}), and $\tilde{I}_{L}(w)$ is given by Eq. (\ref{eq:App-IL-final}).
The details of the derivation of Eq. \eqref{eq:aZ-final} are summarized in Appendix \ref{app:classical_formula}.

\begin{figure}
\begin{centering}
\includegraphics[width=\linewidth]{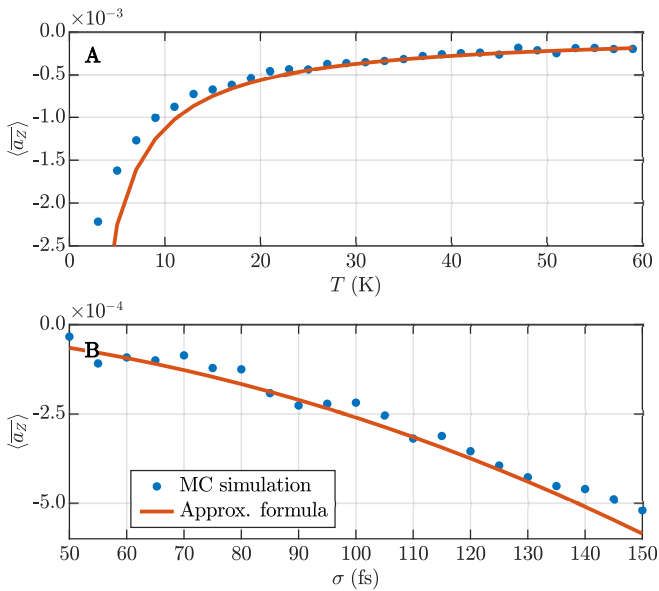}
\par\end{centering}
\caption{(A) Temperature and (B) pulse duration dependence of the degree of long-lasting orientation obtained using the approximate formula in Eq. \eqref{eq:aZ-final} (red line)
and numerically using the impulsive approximation (blue dots). Here $I_\mathrm{FW}=2\times10^{13}\,\mathrm{W/cm^{2}}$,
$I_\mathrm{SH}=0.75\times10^{13}\,\mathrm{W/cm^{2}}$, and $\sigma=120\,\mathrm{fs}$. Each point is an average
of $10^{8}$ sample molecules. In \textbf{B}, the temperature is fixed to $T=30\,\mathrm{K}$.
\label{fig:Fig4}}
\end{figure}

According to Eq. \eqref{eq:aZ-final}, the degree of long-lasting orientation scales as $\sigma^2/T$. In Fig. \ref{fig:Fig4}, we compare the temperature (panel \textbf{A}) and pulse duration (panel \textbf{B}) dependencies of the long-lasting orientation obtained using the approximate formula in Eq. \eqref{eq:aZ-final} with the numerical results obtained by evaluating the formula in Eq. (\ref{eq:abar-i}) using the Monte Carlo approach as described in the Methods Section \ref{sec:Numerical methods} (using the impulsive approximation, see Appendix \ref{app:torques}).

There is a good agreement between the numerical results and the results obtained using the approximate formula, especially at higher temperatures (higher initial angular momenta), where the assumption $|\delta L_{b}/L_{b}|,|\delta L_{c}/L_{c}|\ll1$ is well satisfied. The pulse duration dependence shows the connection to  the energy gained by the molecule from the laser pulse. In the limit of weak excitation (low pulse intensity and/or high temperature), the approximate formula also reveals the more involved dependence on the fields' amplitudes, according to Eqs. \eqref{eq:text-P1} and \eqref{eq:text-P2}.

\begin{figure}[!t]
\centering{}\includegraphics[width=\linewidth]{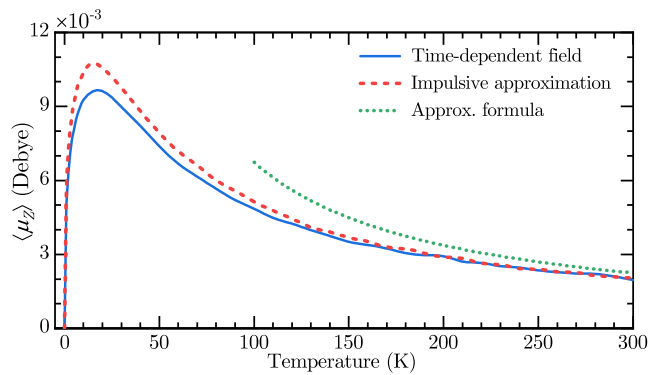}
\caption{Classically calculated permanent values of $Z$-projection of the dipole moment.
The field parameters are similar to Fig. \ref{fig:Fig1}.
Cases of fully time-dependent field (solid blue) and using impulsive approximation (dashed red) are compared. The dotted green line is obtained using the approximate formula in Eq. \eqref{eq:aZ-final}
with $\braket{\mu_Z}=-\mu \langle\overline{a_{Z}}\rangle$ (dipole moment points againts $a$ axis, see Table \ref{tab:Molecular-properties-alternative}).
 \label{fig:temperature_dependence}}
\end{figure}

The hyperpolarizability part of the interaction potential, $\overline{V}_\beta^\star(\theta)$ [see Eq. \eqref{eq:potential-hyperpolarizability2}], is an asymmetric function of $\theta$ (about $\theta=\pi/2$, see Fig. \ref{fig:Fig2}), similar to  the orienting potential, which is proportional to $-\cos(\theta)$, due to a single THz pulse interacting with the molecular dipole, $\boldsymbol{\mu}$. As we show here and as it was shown in \cite{Xu2020}, excitation by such orienting potentials results in transient orientation followed by residual long-lasting orientation.

Despite the similarity, the mechanisms behind the long-lasting orientation induced by a femtosecond two-color and a picosecond THz pulse are not the same. It was shown in \cite{Xu2020} that in the limit of vanishing THz pulse duration, the induced long-lasting orientation tends to zero. In other words, a $\delta$-kick by a purely orienting potential doesn't lead to long-lasting orientation. In contrast, here we show that a $\delta$-kick by a \emph{combined, aligning and orienting}, potentials results in a long-lasting orientation [see the discussion under Eq. \eqref{eq:aZ-integral-2}, also see Fig. \ref{fig:temperature_dependence}].

\section{Temperature and polarization dependence of the long-lasting orientation} \label{sec:SecV-additional-results}

Figure \ref{fig:temperature_dependence} depicts the long-lasting orientation of the dipole moment as a function of temperature for the case of collinearly polarized two-color pulse.
Due to the short pulse duration ($\sigma=120\,\mathrm{fs}$), the results of the fully time-dependent simulation (solid blue line) are well reproduced using the impulsive approximation (dashed red line). The impulsive approximation is described in Appendix \ref{app:torques}. As mentioned in Sec. \ref{sec:SecIII-Qualitative-description}, the long-lasting orientation vanishes at $T=0\,\mathrm{K}$,  [see Eq. \eqref{eq:abar-i}]. At high temperatures, the long-lasting orientation decreases as $\propto T^{-1}$. Therefore, there should be  an optimal temperature for which the long-lasting orientation is maximal.
As is shown in Fig. \ref{fig:temperature_dependence}, for the field parameters used here, the optimal temperature is $T\approx20\,\mathrm{K}$.
In the derivation leading to the approximate formula in Eq. \eqref{eq:aZ-final}, we assumed
$\beta_{bbb}=\beta_{bcc}=0$ . Nevertheless, the results obtained using Eq. \eqref{eq:aZ-final} qualitatively agree with the numerical results at higher temperatures as well (dotted green line).

\begin{figure}[!t]
\centering{}\includegraphics[width=\linewidth]{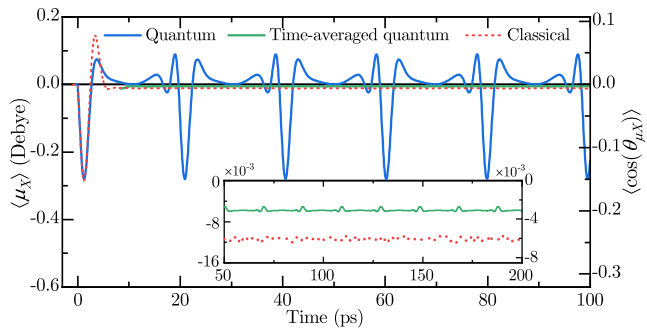}
\caption{$X$-projection of the dipole moment, $\braket{\mu_{X}}$ and the orientation factor, $\braket{\cos(\theta_{\mu X})}\equiv \braket{\mu_{X}}/\mu$ as a function of time.
$\theta_{\mu X}$ denotes the angle between the dipole moment and laboratory
$X$ axis.
The conditions used here are the same as in the case shown in Fig. \ref{fig:Fig1}, except that the angle between the polarizations of FW and SH is $\phi_\mathrm{SH} = \pi/2$ [see Eq. \eqref{eq:electric-field-lab-frame}].
%The solid green line is the time average defined by $\overline{\langle\mu_{Y}\rangle(t)}=(\Delta t)^{-1}\int_{t-\Delta t/2}^{t+\Delta t/2}\mathrm{d}t'\langle\mu_{Y}\rangle(t')$, where $\Delta t=19.6 \,\mathrm{ps}$. The insets show a magnified portion of the signals.
 \label{fig:CH3F_perpendicular}}
\end{figure}

As an additional example, we consider the case of a cross-polarized two-color pulse in which the polarizations of the FW and SH are along $Z$ and $X$ axes, respectively.
Figure \ref{fig:CH3F_perpendicular} shows the dipole signal along the laboratory $X$ axis.
Note that the $Y$ and $Z$-projections of the dipole moment, $\braket{\mu_{Y,Z}}$ are exactly zero.  In this case, a similar transient dipole response along the polarization direction of SH can be seen. On the long time scale, it is followed by the long-lasting orientation. Notice that for the field parameters used here, the achieved degree of both transient and long-lasting orientation is higher in the case of cross-polarized FW and SH compared with Fig. \ref{fig:Fig1} (also see \citep{Mun2019}).

\section{Conclusions} \label{SecVI-Conclusions}

We have theoretically demonstrated a new phenomenon of long-lasting (persistent) orientation of symmetric-top molecules excited by a single two-color femtosecond pulse. The residual orientation was shown to last indefinitely (within the adopted model), or until destroyed by other physical effects, e.g. intermolecular collisions. We derived an approximate classical expression revealing several qualitative features of the phenomenon, including the scaling with temperature, pulse duration, and other field and molecular parameters. The predictions of the formula are in full agreement with the results of numerical simulations in the limit of weak excitation. A quick check for different polarizations showed that in the case of cross-polarized FW and SH, the achieved degree of both transient and long-lasting orientation may be higher than in the case of parallel configuration. Further and careful optimization of  the parameters of the two-color pulse may give rise to even  higher degrees of long-lasting orientation. Moreover, it has been demonstrated  before that the degree of molecular alignment/orientation can be enhanced when a  sequence of several laser pulses is used instead of a single pulse \citep{Averbukh2001,Averbukh2003,Averbukh2004,Lee2004,Bisgaard2004,Pinkham2007,Zhang2011}, and a similar approach may be beneficial for increasing the degree of the long-lasting orientation. The orientation may be measured with the help of second (or higher-order) harmonic generation \cite{Frumker2012} and may be utilized in deflection experiments using inhomogeneous electrostatic fields \cite{Gershnabel2011Electric,*Gershnabel2011Deflection,Kupper2012}.

\appendix

\section{Quantum-mechanical simulation} \label{app:quantum simulation}

We use free symmetric-top wave functions $|JKM\rangle$, the eigenfunctions of kinetic energy Hamiltonian, $H_r$ \citep{zare1988Angular}, as the basis set for numerical calculations. In this basis, the rotational kinetic energy of a prolate top is given by \citep{zare1988Angular}
\begin{align}\label{eq:HR}
\begin{split}
\langle JKM|H_{r}|JKM\rangle=C  J(J+1)+(A-C)K^{2},
\end{split}
\end{align}
where the rotational constants are $C=\hbar^{2}/2I_{b}=\hbar^{2}/2I_{c}$, $A=\hbar^{2}/2I_{a}$ with $A>C$. %$B=C=\hbar^{2}/2I_{c}$ with $A>B=C$.
Using a spherical basis, the transformation between tensors in laboratory- and molecule-fixed reference frames is given by \citep{zare1988Angular}
\begin{align}
T_{p}^{(k)}=\sum\limits _{q}{D_{pq}^{k}}^{*}(R)T_{q}^{(k)},\label{eq:App-spherical-transform}
\end{align}
where ${D_{pq}^{k}}^{*}(R)$ represents the complex conjugate of the Wigner
D-matrix, and $R$ is the set of three Euler angles.
$T_{p}^{(k)}$ and $T_{q}^{(k)}$ denote the spherical tensors of rank $k$ in the laboratory- and molecule-fixed reference frames,
respectively.
Using the transformation in Eq. \eqref{eq:App-spherical-transform}, $V_{\alpha}$ [see Eq. \eqref{eq:potential}] (averaged over the optical cycle) can be expressed as \cite{Yachmenev2016Detecting, Tutunnikov2019Laser}
\begin{align}
\overline{V}_{\alpha}&=
-\sum_{i=1}^2\frac{|\varepsilon_i(t)|^{2}}{4}
\left[
-\frac{1}{\sqrt{3}} \alpha_{\mathrm{mol}, 0}^{(0)} {D_{0,0}^{0}}^*
-\frac{1}{\sqrt{6}} {D_{0, n}^{2}}^* \alpha_{\mathrm{mol}, n}^{(2)}\right. \nonumber
\\ &\left. +\frac{\alpha_{\mathrm{mol}, n}^{(2)}}{2}
\left(e^{i 2 \gamma_i} {D_{-2, n}^{2}}^*+e^{-i 2 \gamma_i} {D_{2, n}^{2}}^*\right)
\right],
\end{align}
where $\gamma$ is the angle between the instantaneous field polarization and the $Z$ axis (according to Eq. \eqref{eq:electric-field-lab-frame}, $\gamma_1 =0$ and $\gamma_2 = \phi_\mathrm{SH} $). The repeated index $n = -2,\cdots,2$ is summed over.
The spherical tensor elements of the molecular polarizability are given by
\begin{align}
\begin{split}
\alpha_{\mathrm{mol}, 0}^{(0)} &=-\frac{1}{\sqrt{3}}\left(\alpha_{a a}+\alpha_{b b}+\alpha_{c c}\right), \\
\alpha_{\mathrm{mol}, 0}^{(2)} &=\frac{1}{\sqrt{6}}\left(2 \alpha_{a a}-\alpha_{b b}-\alpha_{c c}\right), \\
\alpha_{\mathrm{mol},\pm 1}^{(2)} &=\mp \alpha_{a b}-i \alpha_{a c}, \\
\alpha_{\mathrm{mol},\pm 2}^{(2)} &=\frac{1}{2}\left(\alpha_{b b}-\alpha_{c c} \pm 2 i \alpha_{b c}\right).
\end{split}
\end{align}

In addition, $V_{\beta}$ in Eq. \eqref{eq:potential} (averaged over the optical cycle) can be recast into
\begin{equation}\label{eq:potential-spherical-basis}
\overline{V}_{\beta} = -\frac{1}{6} \sum_{j = 0}^3 \sum_{j_{12} = 0}^2 \sum_{m = -j}^{j} (-1)^m  \braket{A_{\mathrm{lab}, m}^{(j, j_{12})}} \beta_{\mathrm{lab}, -m}^{(j, j_{12})},
\end{equation}
where $A_{\mathrm{lab}, m}^{(j, j_{12})}$ and $\beta_{\mathrm{lab}, m}^{(j, j_{12})}$
are the spherical tensor elements of the electric field and the molecular hyperpolarizability in the laboratory reference frame, respectively.
The spherical tensor $A_{\mathrm{lab}}^{(j_3, j_{12})}$ can be constructed by performing dyadic product of $A^{(1)}$ twice \cite{Man2013Cartesian, devanathan2002vectors}, namely
\begin{small}\begin{align}
\begin{split}
A_{m_{12}}^{(j_{12})}&=\sum_{m_1 = -1}^1\sum_{m_2 = -1}^1
\braket{1 m_1 1 m_2
|j_{12}m_{12}} A_{m_{1}}^{(1)} A_{m_{2}}^{(1)},\\
A_{\mathrm{lab}, m}^{(j, j_{12})}&=\sum_{m_{12} = -j_{12}}^{j_{12}}\sum_{m_3 = -1}^1
\braket{j_{12} m_{12} 1 m_3 |j m} A_{m_{12}}^{(j_{12})} A_{m_{3}}^{(1)},
\end{split}
\end{align}
\end{small}
where $\braket{1 m_1 1 m_2 |j_{12}m_{12}}$ and $\braket{j_{12} m_{12} 1 m_3 |j m}$ represent the Clebsch-Gordan coefficients. $A^{(1)}$ is the electric field's spherical tensor of rank one and its components are
given by $A_{\pm1}^{(1)}=\mp(E_{Z}\pm iE_{X})/\sqrt{2}$ and $A_{0}^{(1)}=E_{Y}$.
For the two-color field defined in Eq. \eqref{eq:electric-field-lab-frame}, the nonzero components of $\braket{A_{\mathrm{lab}, m}^{(j, j_{12})}}$ are
\begin{align}
\begin{split}
\braket{A_{\mathrm{lab}, \pm1}^{(1, 0)}} &= \pm\frac{\varepsilon_1^2 \varepsilon_2}{4\sqrt{6}}[3\cos(\phi_{\mathrm{SH}}) \pm i \sin(\phi_{\mathrm{SH}})],\\
\braket{A_{\mathrm{lab}, \pm1}^{(1, 2)}} &= \pm\frac{\varepsilon_1^2 \varepsilon_2}{2\sqrt{30}} [3\cos(\phi_{\mathrm{SH}}) \pm i \sin(\phi_{\mathrm{SH}})],\\
\braket{A_{\mathrm{lab}, \pm1}^{(3, 2)}} &= \pm\frac{3\varepsilon_1^2 \varepsilon_2}{8\sqrt{30}} [ 3\cos(\phi_{\mathrm{SH}}) \pm i \sin(\phi_{\mathrm{SH}})],\\
\braket{A_{\mathrm{lab}, \pm3}^{(3, 2)}} &= \mp\frac{\varepsilon_1^2 \varepsilon_2}{8\sqrt{2}} [\cos(\phi_{\mathrm{SH}}) \pm i \sin(\phi_{\mathrm{SH}})],
\end{split}
\end{align}
while the hyperpolarizability in the laboratory reference frame is given by
$\beta_{\mathrm{lab}, p}^{(j, j_{12})} = \sum\limits _{q}{D_{pq}^{j}}^{*}(R)\beta_{\mathrm{mol}, q}^{(j, j_{12})}$, where
{\small{}
\begin{align}
\begin{split}
    \beta_{\mathrm{mol}, 0}^{(1, 0)}        =& -\frac{1}{\sqrt{3}}
(\beta_{aaa}+\beta_{abb}+\beta_{acc}) ,\\
    \beta_{\mathrm{mol}, \pm1}^{(1, 0)}     =& \pm\frac{1}{\sqrt{6}}
\Big[ (\beta_{aab}+\beta_{bbb}+\beta_{bcc}) \\
& \pm i(\beta_{aac}+\beta_{bbc}+\beta_{ccc})\Big],\\
    \beta_{\mathrm{mol}, 0}^{(1, 2)}        =& -\frac{2}{\sqrt{15}}
(\beta_{aaa}+\beta_{abb}+\beta_{acc}),\\
    \beta_{\mathrm{mol}, \pm1}^{(1, 2)}     =& \pm\frac{2}{\sqrt{30}}
\Big[(\beta_{aab}+\beta_{bbb}+\beta_{bcc})  \\
& \pm i (\beta_{aac}+\beta_{bbc}+\beta_{ccc})\Big],\\
    \beta_{\mathrm{mol}, 0}^{(3, 2)}        =& -\frac{1}{\sqrt{10}}
(3\beta_{abb}+3\beta_{acc}-2\beta_{aaa}),\\
    \beta_{\mathrm{mol}, \pm1}^{(3, 2)}     =& \pm \frac{3}{2\sqrt{30}} %\frac{1}{2}\sqrt{\frac{3}{10}}
\Big[ (\beta_{bbb}+\beta_{bcc}-4\beta_{aab}) \\
& \pm i(\beta_{bbc}+\beta_{ccc}-4\beta_{aac})\Big],\\
    \beta_{\mathrm{mol}, \pm2}^{(3, 2)}     =& \frac{\sqrt{3}}{2}
(\beta_{abb} - \beta_{acc} \pm 2i\beta_{abc}),\\
    \beta_{\mathrm{mol}, \pm3}^{(3, 2)}     =& \mp\frac{1}{2\sqrt{2}}
\Big[ (\beta_{bbb}-3\beta_{bcc}) \mp i(-3\beta_{bbc}+\beta_{ccc})\Big].
\end{split}
\end{align}
}

The matrix elements of the molecule-field interaction potential, $\langle JKM|H_{\mathrm{int}}|J'K'M'\rangle$ can be obtained with the help of the relation \citep{zare1988Angular}
\begin{align}
&\quad \langle JKM |{D_{pq}^{s}}^{*}(R)|J'K'M'\rangle=(-1)^{M-K} \times \nonumber\\
&\sqrt{(2J+1)(2J'+1)} \begin{pmatrix}J & s & J'\\
-M & p & M'
\end{pmatrix}\begin{pmatrix}J & s & J'\\
-K & q & K'
\end{pmatrix}, \!\!
\end{align}
where the large brackets represent the Wigner 3-$j$ symbols.

The polarization [Eq. \eqref{eq:Polarization}] reads
\begin{align}\label{eq:Polarization2}
 &\braket{\mu_i^{(J,K,M)}}\!(t) \nonumber\\
 =&\sum\limits _{p,q=-1}^{+1}(-1)^{p}\mu_{q}^{(1)}e_{-p}^{(1)}\langle\Psi(t)|{D_{pq}^{1}}^{*}(R)|\Psi(t)\rangle,
\end{align}
where $e_{\pm1}^{(1)}=\mp(e_{Z}\pm ie_{X})/\sqrt{2}$ and $e_{0}^{(1)}=e_{Y}$.
The components of spherical tensor representing the dipole moment in the molecular reference frame are defined by
by $\mu_{\pm1}^{(1)}=\mp(\mu_{b}\pm i\mu_{c})/\sqrt{2}$ and $\mu_{0}^{(1)}=\mu_{a}$.

\section{Derivation of formulas in Eqs. (\ref{eq:delta-Lb}, \ref{eq:delta-Lc})}
\label{app:torques}

The Lagrangian of the classical symmetric-top molecule excited by a two-color pulse reads
\begin{align}
\mathcal{L}  =\frac{L_a^2}{2I_a}+ \frac{L_{b}^2+L_{c}^2}{2I} - \overline{V}(\theta,t),\label{eq:Lagrangian}
\end{align}
where the set of canonical variables includes the three Euler angles and their time derivatives, $(\theta, \phi, \chi; \dot{\theta}, \dot{\phi},\dot{\chi})$. The interaction potential $\overline{V}(\theta,t)$ is given by Eqs. \eqref{eq:potential-polarizability} and \eqref{eq:potential-hyperpolarizability2}.
In terms of the canonical variables, the components of angular momentum (in the basis of molecular principal axes of inertia) are given by \citep{zare1988Angular}
\begin{align}
\begin{split}
&L_b  = I \left[-\dot{\phi}\sin(\theta)\cos(\chi)+\dot{\theta}\sin(\chi)\right],\\
&L_c = I \left[\dot{\phi}\sin(\theta)\sin(\chi)+\dot{\theta}\cos(\chi)\right],\\
&L_a  =I_a  \left[ \dot{\phi}\cos(\theta)+\dot{\chi}\right],
\end{split}
\end{align}
such that the Lagrangian in Eq. \eqref{eq:Lagrangian} becomes
\begin{align}
\mathcal{L} & = \frac{I_a}{2}\left[\dot{\phi}\cos(\theta)+\dot{\chi}\right]^2\nonumber\\
&+\frac{I}{2}\left[\dot{\phi}^2\sin^2(\theta)+\dot{\theta}^2\right]- \overline{V}(\theta,t).
\end{align}

Assuming the angles remain constant during the impulsive excitation, the changes of angular momentum components are given by
\begin{align}
\begin{split}
\delta L_b
&= I \left[-\delta\dot{\phi}\sin(\theta)\cos(\chi)+\delta\dot{\theta}\sin(\chi)\right]
,\\
\delta L_c
&= I \left[\delta\dot{\phi}\sin(\theta)\sin(\chi)+\delta\dot{\theta}\cos(\chi)\right]
,\\
\delta L_a
&=I_a  \left[ \delta\dot{\phi}\cos(\theta)+\delta\dot{\chi}\right],
\end{split}
\end{align}
where $\delta g = g(t_f) - g(t_i)$ denotes the difference between a quantity after ($t_{f}$) and before ($t_{i}$) the pulse. The changes of the first derivatives, $\delta\dot{\theta}$, $\delta\dot{\phi}$, $\delta\dot{\chi}$ can be obtained by integrating the Euler-Lagrange equations
\begin{fleqn}
\begin{align}
&\frac{d}{dt}\left(\frac{\partial \mathcal{L}}{\partial{\dot{\theta}}}\right)=\frac{d}{dt}\left(I\dot{\theta}\right) =\frac{\partial \mathcal{L} }{\partial {\theta}}\nonumber\\=&  -I_a \left[\dot{\phi}\cos(\theta)+\dot{\chi}\right] \sin(\theta)+I\dot{\phi}^2\sin(\theta)\cos(\theta)-\frac{\partial \overline{V}(\theta)}{\partial {\theta}},\nonumber\\
&\frac{d}{dt}\left(\frac{\partial \mathcal{L}}{\partial{\dot{\phi}}}\right)=\frac{d}{dt}\left(I_a\left[\dot{\phi}\cos(\theta)+\dot{\chi}\right]\cos(\theta)+I\dot{\phi}\sin^2(\theta)\right) \nonumber\\
= &\frac{\partial \mathcal{L} }{\partial {\phi}}=0,\nonumber\\
&\frac{d}{dt}\left(\frac{\partial \mathcal{L}}{\partial{\dot{\chi}}}\right)=\frac{d}{dt}\left(I_a\left[\dot{\phi}\cos(\theta)+\dot{\chi}\right]\right)  = \frac{\partial\mathcal{L} }{\partial {\chi}}=0.
\end{align}
\end{fleqn}
once and solving the resulting system of equations for $\delta\dot{\theta}$, $\delta\dot{\phi}$, $\delta\dot{\chi}$. During the integration, the angles are assumed to be constant, so that integrals such as $\int_{t_i}^{t_f}\dot{\phi}\,dt, \int_{t_i}^{t_f}\dot{\chi}\,dt, \int_{t_i}^{t_f}\dot{\phi}^2\,dt$, etc. vanish. Overall, we end up with
\begin{align}
\delta{\dot{\phi}}=\delta{\dot{\chi}} =0,\quad
\delta{\dot{\theta}} = -\frac{1}{I}\int_{t_i}^{t_f}\frac{\partial \overline{V}(\theta,t)}{\partial {\theta}} d t,
\end{align}
such that
\begin{align}
\begin{split}
\delta L_b
&= -\sin(\chi)\int_{t_i}^{t_f}\frac{\partial \overline{V}(\theta,t)}{\partial {\theta}} d t
,\\
\delta L_c
&= -\cos(\chi)\int_{t_i}^{t_f}\frac{\partial \overline{V}(\theta,t)}{\partial {\theta}} d t
,\\
\delta L_a
& = 0.
\end{split}
\end{align}

\section{Derivation of formula in Eq. \eqref{eq:aZ-final}}
\label{app:classical_formula}

\begin{figure}[t]
\begin{centering}
\includegraphics[width=\linewidth]{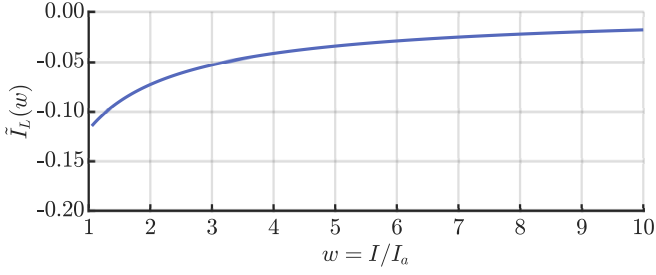}
\par\end{centering}
\caption{Graph of the function $\tilde{I}_L(w)$, see Eq. \eqref{eq:App-IL-final}.
\label{fig:app-Fig1}}
\end{figure}

In this section, we carry out the intermediate steps in the derivation
of Eqs. (\ref{eq:aZ-integral-2}) and (\ref{eq:aZ-final}). After $d\phi$ integration, Eq. (\ref{eq:aZ-integral-1}) becomes
\begin{align}
\langle\overline{a_{Z}}\rangle & =\frac{2\pi}{\mathcal{Z}}\int_{L^3}\iint\frac{L_{Z}L_{a}}{L_{f}^{2}}\nonumber \\
&\times\exp\left[-\frac{1}{2k_{B}T}\left(\frac{L_{a}^{2}}{I_{a}}+\frac{L_{b}^{2}}{I}+\frac{L_{c}^{2}}{I}\right)\right]\nonumber \\
 & \times\sin\theta\,d\theta d\chi dL_{a}dL_{b}dL_{c}.\label{eq:App-aZ-integral-1}
\end{align}
Next, we expand $1/L_{f}^{2}$ in powers of $f(\theta)$ and carry
out $d\chi$ integration. Only terms proportional to even powers
of $f(\theta)$ survive the integration. The first non-vanishing term
is proportional to $f^{2}(\theta)$, such that
\begin{align}
\langle\overline{a_{Z}}\rangle & \approx\frac{2\pi^{2}}{\mathcal{Z}}I_{L}\int f^{2}(\theta)\sin2\theta\,d\theta,\label{eq:App-aZ-integral-2}
\end{align}
where
\begin{fleqn}
\begin{align}
I_{L} & =\int_{L^3}\frac{L_{a}^{2}(L_{b}^{2}+L_{c}^{2}-L_{a}^{2})}{L^{6}}\nonumber \\
 &\times\exp\left[-\frac{1}{2k_{B}T}\left(\frac{L_{a}^{2}}{I_{a}}+\frac{L_{b}^{2}}{I}+\frac{L_{c}^{2}}{I}\right)\right]dL_{a}dL_{b}dL_{c}.\!\!\!\!\label{eq:App-IL-integral-1}
\end{align}
\end{fleqn}
The temperature dependence is isolated in $I_{L}$ and can be exposed
by change of variables. For this, we substitute
$L_{b}^{2}+L_{c}^{2}=L^{2}-L_{a}^{2}$, such that
\begin{fleqn}\begin{align}
I_{L} & =\int_{L^3}\frac{L_{a}^{2}(L^{2}-2L_{a}^{2})}{L^{6}}\nonumber \\
 &\times \exp\left[-\frac{L^{2}}{2k_{B}TI}\left(1+\frac{L_{a}^{2}}{L^{2}}(w-1)\right)\right]\,dL_{a}dL_{b}dL_{c},\!\!\!\!\!\!\!\label{eq:App-IL-integral-2}
\end{align}
\end{fleqn}
where $w=I/I_{a}$. Changing to spherical coordinates, and integrating
$d\phi_{L}$ produces
\begin{align}
I_{L} & =2\pi\iint\cos^{2}(\theta_{L})\sin(\theta_{L})[1-2\cos^{2}(\theta_{L})].\nonumber \\
 & \times\exp\left[-\frac{L^{2}}{2k_{B}TI}[1+\cos^{2}(\theta_{L})(w-1)]\right]\,d\theta_{L}dL.\label{eq:App-IL-integral-3}
\end{align}
Next, we introduce
\[
l=\frac{L}{\sqrt{2k_{B}TI}},
\]
such that
\begin{align}
I_{L}(w) & =2\pi\sqrt{2k_{B}TI}\iint\cos^{2}(\theta_{L})\sin(\theta_{L})[1-2\cos^{2}(\theta_{L})]\nonumber \\
 & \times\exp\left[-l^{2}[1+\cos^{2}(\theta_{L})(w-1)]\right]\,d\theta_{L}dl.\label{eq:App-IL-integral-4}
\end{align}
For the partition function, $\mathcal{Z}$, we have
\begin{align}
\mathcal{Z} & =(2k_{B}TI)^{3/2}\iint\sin(\theta)\,d\theta d\chi d\phi\nonumber \\
 & \times\int_{V_l} l^{2}\exp\left[-l^{2}[1+\cos^{2}(\theta_{L})(w-1)]\right]\nonumber \\
&\times\sin(\theta_{L})\,d\phi_{L}d\theta_{L}dl=16\pi^{3}(2k_{B}TI)^{3/2}I_{\mathcal{Z}}(w),\label{eq:App-Z-integral}
\end{align}
where
\begin{align}
I_{\mathcal{Z}}(w) & =\iint l^{2}\exp\left[-l^{2}[1+\cos^{2}(\theta_{L})(w-1)]\right]\nonumber \\
&\times\sin(\theta_{L})\,d\theta_{L}dl=\frac{\sqrt{\pi}}{2\sqrt{w}}.\label{eq:App-IZ-integral}
\end{align}
Thus,
\begin{align}
\langle\overline{a_{Z}}\rangle & \approx\frac{4\pi^{3}\sqrt{2k_{B}TI}}{16\pi^{3}(2k_{B}TI)^{3/2}}\frac{\tilde{I}_{L}(w)}{I_{\mathcal{Z}}(w)}\int f^{2}(\theta)\sin2\theta\,d\theta\nonumber \\
 & =\frac{\tilde{I}_{L}(w)}{4k_{B}TI}\sqrt{\frac{w}{\pi}}\int f^{2}(\theta)\sin2\theta\,d\theta,
% & =\frac{1}{8k_{B}TI}\frac{\tilde{I}_{L}(w)}{I_{\mathcal{Z}}}\int f^{2}(\theta)\sin2\theta\,d\theta,
 \label{eq:App-aZ-integral-3}
\end{align}
where
\begin{align}
\tilde{I}_{L}(w) &=\frac{I_{L}(w)}{2\pi\sqrt{2k_{B}TI}}\nonumber \\
 =&\frac{\sqrt{\pi}[3\sqrt{(w-1)w}-(2w+1)\cosh^{-1}(\sqrt{w})]}{4(w-1)^{5/2}},\label{eq:App-IL-final}
\end{align}
see Fig. \ref{fig:app-Fig1}.
When both polarizability and hyperpolarizability interactions are
included, $f(\theta)$ is given by Eq. (\ref{eq:text-f(theta)}),
and
\[
\int f^{2}(\theta)\sin2\theta\,d\theta=\frac{64}{105}P_{1}P_{2}\left(2\beta_{abb}-3\beta_{aaa}\right).
\]
Overall, we have
\begin{fleqn}\begin{equation}
\langle\overline{a_{Z}}\rangle=\frac{16\tilde{I}_{L}(w)}{105k_{B}TI}\sqrt{\frac{w}{\pi}}P_{1}P_{2}\left(2\beta_{abb}-3\beta_{aaa}\right),\label{eq:App-aZ-final}
\end{equation}
\end{fleqn}
where $P_{1}$ and $P_{2}$ are given by Eqs. (\ref{eq:text-P1})
and (\ref{eq:text-P2}).\\

\section*{Conflict of Interest Statement}
The authors declare that the research was conducted in the absence of any commercial or financial relationships that could be construed as a potential conflict of interest.

\section*{Data Availability Statement}
The raw data supporting the conclusions of this article will be
made available by the authors, without undue reservation.

\section*{Author Contributions}
All the authors participated in formulating the problem and initiating this study.   L.X. and  I.T. equally contributed to the calculations and numerical simulations. All the authors participated in analyzing the results and writing the manuscript. Y.P. and I.A. supervised and guided the work.

\section*{Funding}
We gratefully acknowledge support by the Israel Science
Foundation (Grant No. 746/15). I.A. acknowledges support as the Patricia Elman
Bildner Professorial Chair. This research was made possible in
part by the historic generosity of the Harold Perlman Family.

\end{document}